# Ransomware Detection and Classification Using Random Forest: A Case Study with the UGRansome2024 Dataset

**Peace Azugo,** u21485730@tuks.co.za, *Department of Computer Science, University of Pretoria, South Africa*
**Hein Venter,** hventer@cs.up.ac.za, *Department of Computer Science, University of Pretoria, South Africa*
**Mike Wa Nkongolo,** mike.wankongolo@up.ac.za, *Department of Informatics, University of Pretoria, South Africa*

**Abstract**

Cybersecurity faces challenges in identifying and mitigating ransomware, which is important for protecting critical infrastructures. The absence of datasets for distinguishing normal versus abnormal network behaviour hinders the development of proactive detection strategies against ransomware. An obstacle in proactive prevention methods is the absence of comprehensive datasets for contrasting normal versus abnormal network behaviours. The dataset enabling such contrasts would significantly expedite threat anomaly mitigation. In this study, we introduce *UGRansome2024*, an optimised dataset for ransomware detection in network traffic. This dataset is derived from the *UGRansome* data using an intuitionistic feature engineering approach that considers only relevant patterns in network behaviour analysis. The study presents an analysis of ransomware detection using the UGRansome2024 dataset and the Random Forest algorithm. Through encoding and feature relevance determination, the Random Forest achieved a classification accuracy of 96% and effectively identified unusual ransomware transactions. Findings indicate that certain ransomware variants, such as those utilising *Encrypt Decrypt Algorithms* (EDA) and *Globe* ransomware, have the highest financial impact. These insights have significant implications for real-world cybersecurity practices, highlighting the importance of machine learning in ransomware detection and mitigation. Further research is recommended to expand datasets, explore alternative detection methods, and address limitations in current approaches.

1. **INTRODUCTION**

Ransomware has emerged as a significant cybersecurity threat, posing grave risks to critical infrastructure worldwide [1]. Ransomware's programs encrypt users' files or systems and demand payment for their release, often causing disruptions in critical services and substantial financial losses [2]. For example, the WannaCry ransomware attack in 2017 targeted computers running Microsoft Windows operating systems, affecting organisations such as the UK's National Health Service (NHS) and disrupting healthcare services. Despite the increasing prevalence of ransomware attacks, detecting and mitigating them remains a formidable challenge [1, 2, 3]. One of the primary obstacles is the lack of comprehensive datasets for training machine learning models to identify ransomware behaviour accurately [4]. Without access to large and diverse datasets containing both normal and ransomware network traffic, developing effective detection strategies becomes exceedingly difficult [4, 5]. To address this challenge, the UGRansome dataset was introduced in 2021 as a resource for detecting ransomware in network traffic [1, 2, 3, 5]. However, since its inception, the dataset has not been optimised with new features to keep pace with the evolving landscape of ransomware threats. With new ransomware variants and network attacks emerging regularly, the static nature of the UGRansome dataset hinders its effectiveness in detecting novel ransomware strains. This research aims to enhance the UGRansome dataset by incorporating new features and optimising its performance for ransomware detection. Specifically, the dataset has been updated to UGRansome2024, integrating intuitive feature engineering methodologies and random sampling techniques to enhance its efficacy [6]. By enriching the dataset with relevant patterns and characteristics of ransomware behaviour, the updated UGRansome2024 dataset offers improved capabilities for identifying and classifying ransomware activities in network traffic. Additionally, the application of the random forest algorithm further enhances the detection accuracy and efficiency of the dataset, making it a valuable resource for cybersecurity professionals and researchers alike. The manuscript is structured into the following sections:

- ***Background***: This section provides context and foundational information related to ransomware detection, emphasising the need for innovative approaches to dataset creation.
- ***Theory-Guided Design Ransomware Dataset***: Here, the concept of *theory-guided design* for ransomware dataset creation is introduced, discussing the theoretical framework guiding the dataset development process [9].
- ***UGRansome Dataset***: This section delves into the specifics of the UGRansome dataset, outlining its features, composition, and significance in the field of ransomware analysis.
- ***Feature Engineering***: The process of feature engineering employed in refining the UGRansome dataset is described, highlighting the techniques used to enhance its effectiveness and utility.
- ***Ransomware Detection:*** Details regarding the application of ransomware detection techniques using the UGRansome dataset are discussed, including methodologies and algorithms utilised.
- ***Random Forest Results***: Results obtained from employing the Random Forest algorithm for ransomware detection using the UGRansome dataset are presented and analysed in this section.
- ***Significance of the Findings***: The significance and implications of the research findings are examined, elucidating their importance in advancing ransomware detection capabilities.
- ***Applicability in Real-Life***: This section explores the practical implications of the research findings and their potential applications in real-world cybersecurity scenarios.
- ***Future Studies***: Suggestions for future research directions and areas of exploration within the realm of ransomware detection and dataset development are outlined.
- ***Limitations***: Potential limitations and challenges encountered during the research process are acknowledged and discussed, providing insights into areas for improvement.
- ***Recommendation***: Recommendations for addressing the identified limitations and further enhancing the effectiveness of ransomware detection methodologies are proposed.
- ***Conclusion***: The manuscript concludes with a summary of the key findings, insights gained, and the overarching significance of the research contributions.

## 2. BACKGROUND

As organisations and individuals increasingly fall victim to ransomware attacks, the need for effective detection and mitigation strategies has become paramount. One essential tool in the fight against ransomware is the availability of comprehensive datasets for training machine learning models and developing detection algorithms [7]. These datasets play a crucial role in understanding the characteristics and behaviours of ransomware, enabling researchers and cybersecurity professionals to devise effective defence mechanisms [8]. However, building a reliable ransomware dataset poses several challenges [7, 8]. Firstly, ransomware attacks are constantly evolving, with new variants and techniques emerging regularly. This dynamic nature makes it difficult to capture and characterise ransomware behaviour accurately [3]. Secondly, ransomware attacks often target sensitive or proprietary information, making it challenging to obtain real-world ransomware samples for analysis without compromising data privacy and security [9]. Despite efforts to create ransomware datasets, existing ones have several limitations [10]. Many datasets are limited in size and scope, containing only a subset of ransomware samples or focusing on specific ransomware families [10, 11]. Additionally, the quality and diversity of samples in these datasets may vary, affecting the robustness and generalizability of detection models trained on them [3, 5, 6, 7, 10, 11]. Given these challenges and limitations (Table 1), there is a pressing need for innovative approaches to ransomware dataset creation. Feature engineering, a process that involves selecting, extracting, and transforming relevant features from raw data, offers a promising solution [9, 12]. By carefully designing and engineering features that capture the unique characteristics of ransomware behaviour, researchers can create new datasets that address the shortcomings of existing ones [9].

| Reference | Limitation | Recommendation |
| --- | --- | --- |

| [3] | Difficulty capturing evolving ransomware behaviour accurately. | Implement dynamic data collection techniques to continuously update datasets with new ransomware samples and behaviours. |
| --- | --- | --- |
| [7] | Limited size and scope, lack of diversity in samples. | Develop larger and more diverse datasets that encompass a wider range of ransomware variants and behaviours. |
| [8] | Limited quality of samples, lack of representative data. | Ensure datasets contain high-quality samples that accurately represent real-world ransomware behaviour. |
| [9] | Challenges in obtaining real-world ransomware samples. | Collaborate with cybersecurity agencies and organisations to access anonymized ransomware samples for analysis. |
| [10] | Limited scope, focusing on specific ransomware families. | Expand dataset coverage to include a broader range of ransomware families and variants. |
| [11] | Lack of comprehensive coverage, missing critical ransomware behaviours. | Conduct thorough analysis to identify and incorporate missing ransomware behaviours into datasets. |
| [12] | Intuitionistic fuzzy Entropy method. | Utilise feature engineering techniques to create new datasets that address the shortcomings of existing ones. |

**Table 1**: Limitation of Existing Studies

In this context, this paper explores the importance of feature engineering in the development of ransomware datasets. We discuss the challenges associated with building ransomware datasets, examine limitations of current datasets, and highlight the need for novel approaches to dataset creation (Table 1). Through feature engineering, we aim to overcome these challenges and contribute to the advancement of ransomware detection and cybersecurity research.

**THEORY GUIDED DESIGN DATASET CREATION**
Theory-guided design dataset creation involves leveraging existing theoretical knowledge or established principles to inform the construction of datasets for machine learning or data analysis tasks [9]. Rather than relying solely on empirical data collection or random sampling, theory-guided design incorporates domain expertise, scientific theories, or established frameworks to guide the selection, extraction, and generation of data features [9]. In the context of ransomware detection, theory-guided design dataset creation may involve incorporating knowledge about ransomware behaviour, attack patterns, and propagation mechanisms into the dataset construction process. This approach ensures that the dataset reflects the underlying principles and characteristics of ransomware, making it more effective for training and evaluating detection algorithms. By integrating theoretical insights into dataset creation, theory-guided design aims to produce more comprehensive, representative, and informative datasets [9]. These datasets are better suited for addressing specific research questions, testing hypotheses, and developing robust machine learning models or analytical techniques in various domains, including cybersecurity, healthcare, finance, and social sciences. Therefore, based on the challenges and recommendations identified in the

Table 1, we propose a *high-level pyramidal framework for theory-guided design* of ransomware datasets creation:

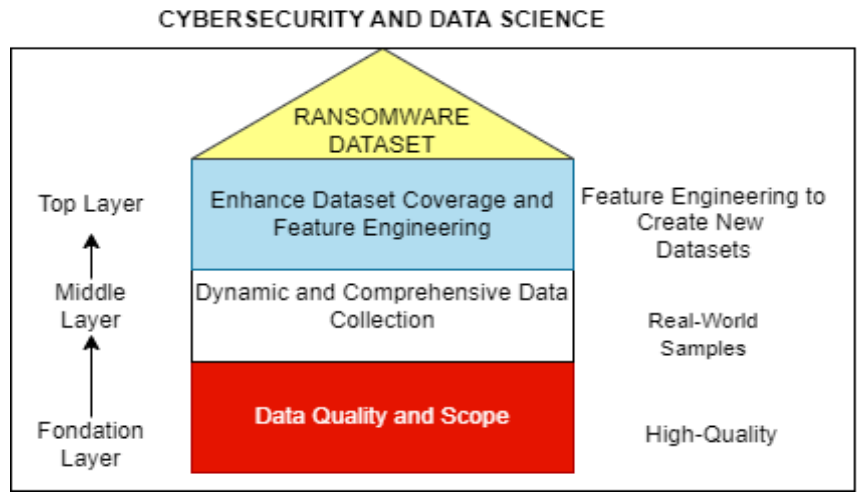

**Figure 1**: Theory Guided Framework for Ransomware Datasets

**Foundation Layer: Addressing Data Quality and Scope**
In this layer, researchers should develop larger and more diverse datasets that encompass a wider range of ransomware variants and behaviours to address the limited size, scope, and lack of diversity in samples [7]. They must ensure datasets contain high-quality samples that accurately represent real-world ransomware behaviour to overcome limitations in sample quality and data representation [8].

**Middle Layer: Dynamic and Comprehensive Data Collection**
In this layer, researchers could implement dynamic data collection techniques to continuously update datasets with new ransomware samples and behaviours by addressing the challenge of capturing evolving ransomware behaviour accurately [3]. They can collaborate with cybersecurity agencies and organisations to access anonymized ransomware samples for analysis and overcome challenges in obtaining real-world ransomware samples [9].

**Top Layer: Enhanced Dataset Coverage and Feature Engineering**
In this layer, researchers must expand dataset coverage to include a broader range of ransomware families and variants, moving beyond specific ransomware families to provide a more comprehensive view of ransomware behaviour [10]. They can conduct thorough analysis to identify and incorporate missing ransomware behaviours into datasets, ensuring comprehensive coverage and addressing any gaps in critical ransomware behaviours [11]. They can also utilise feature engineering techniques to create new datasets that address the shortcomings of existing ones, such as limited scope or missing critical behaviours [12]. The framework illustrated in Figure 1 outlines a systematic approach to theory-guided design of ransomware datasets, starting from foundational improvements in data quality and scope, advancing to dynamic and comprehensive data collection strategies, and culminating in the creation of enhanced datasets through feature engineering and analysis.

**THE UGRANSOME DATASET**
The UGRansome dataset was created in 2021 as a robust cybersecurity resource tailored for the analysis of ransomware and zero-day cyberattacks, particularly those exhibiting cyclostationary behaviour [1]. This dataset encompasses several crucial components essential for comprehensive analysis (Figure 2):

- *Timestamps*: Facilitating attack time tracking for precise temporal analysis.
- *Flags*: Enabling categorization of attack types, aiding in classification and understanding.
- **Protocol Data**: Providing insights into attack vectors, enhancing understanding of attack methodologies.

- **Network Flow Details**: Observing data transfer patterns, crucial for identifying anomalous behaviour.
- **Ransomware Family Classifications**: Categorising ransomware variants, aiding in threat identification and classification.
- **Malware Analysis**: Offering insights into associated malware, aiding in understanding attack payloads.
- **Numeric Clustering**: Utilising pattern recognition techniques to identify common attack patterns.
- **Financial Damage Quantification**: Providing metrics in both USD and bitcoins (BTC) to quantify the impact of attacks.
- **Machine Learning**: Leveraging machine learning algorithms to generate attack signatures, enhancing detection capabilities.
- **Synthetic Signatures**: Offering synthetic signatures for testing and simulating cybersecurity defences, facilitating evaluation and validation.
- **Anomaly Detection**: Enabling the identification and documentation of anomalies, contributing to anomaly detection research.

This dataset has been meticulously curated to offer valuable information for researchers and practitioners alike, supporting various analytical and investigatory purposes, including ransomware and zero-day threats detection and classification [2, 3, 4]. The UGRansome dataset can be downloaded and processed on the Kaggle platform (https://www.kaggle.com/dsv/7172543). This dataset has received various citations from different sources (Table 2).

| | Time | Protocol | Flag | Family | Clusters | SeedAddress | ExpAddress | BTC | USD | Netflow_Bytes | IPaddress | Threats | Port | Prediction |
|---|---|---|---|---|---|---|---|---|---|---|---|---|---|---|
| 0 | 50 | TCP | A | WannaCry | 1 | 1DA11mPS | 1BonuSr7 | 1 | 500 | 5 | A | Botnet | 5061 | SS |
| 1 | 40 | TCP | A | WannaCry | 1 | 1DA11mPS | 1BonuSr7 | 1 | 504 | 8 | A | Botnet | 5061 | SS |
| 2 | 30 | TCP | A | WannaCry | 1 | 1DA11mPS | 1BonuSr7 | 1 | 508 | 7 | A | Botnet | 5061 | SS |
| 3 | 20 | TCP | A | WannaCry | 1 | 1DA11mPS | 1BonuSr7 | 1 | 512 | 15 | A | Botnet | 5061 | SS |
| 4 | 57 | TCP | A | WannaCry | 1 | 1DA11mPS | 1BonuSr7 | 1 | 516 | 9 | A | Botnet | 5061 | SS |
| ... | ... | ... | ... | ... | ... | ... | ... | ... | ... | ... | ... | ... | ... | ... |
| 149038 | 33 | UDP | AP | TowerWeb | 3 | 1AEoiHYZ | 1SYSTEMQ | 1010 | 1590 | 3340 | A | Scan | 5062 | A |
| 149039 | 33 | UDP | AP | TowerWeb | 3 | 1AEoiHYZ | 1SYSTEMQ | 1014 | 1596 | 3351 | A | Scan | 5062 | A |
| 149040 | 33 | UDP | AP | TowerWeb | 3 | 1AEoiHYZ | 1SYSTEMQ | 1018 | 1602 | 3362 | A | Scan | 5062 | A |
| 149041 | 33 | UDP | AP | TowerWeb | 3 | 1AEoiHYZ | 1SYSTEMQ | 1022 | 1608 | 3373 | A | Scan | 5062 | A |
| 149042 | 33 | UDP | AP | TowerWeb | 3 | 1AEoiHYZ | 1SYSTEMQ | 1026 | 1614 | 3384 | A | Scan | 5062 | A |

149043 rows × 14 columns

**Figure 2**: The UGRansome dataset

| Reference | Aim | Limitation |
|---|---|---|
| [13] | To improve ransomware attack detection using transfer learning and deep learning ensemble models on cloud-encrypted data. | Dependency on cloud infrastructure for data encryption may not be feasible in all scenarios. |
| [14] | To create a modern netflow network dataset with labelled attacks and detection methods for ransomware detection. | Limited diversity in attack scenarios and behaviours covered in the dataset. |
| [15] | To conduct a deep analysis of risks and recent trends towards network intrusion detection systems, including ransomware | Limited focus specifically on ransomware detection methods and techniques. |
|  | detection. |  |
| [16] | To propose a signature-based botnet (emotet) detection mechanism, which can indirectly contribute to ransomware detection. | Specific focus on detecting emotet botnet, may not generalise to all ransomware detection scenarios. |
| [17] | To develop a deep forest approach for zero-day attacks detection, which can potentially be applied to ransomware detection. | Limited evaluation of the deep forest approach in real-world ransomware detection scenarios. |
| [18] | To propose strategies for mitigating cybersecurity risks in the US healthcare sector, including ransomware detection and prevention measures. | Focus on cybersecurity risks in a specific sector may limit generalizability to other industries. |
| [19] | To introduce CESSO-HCRNN, a hybrid CRNN with chaotic enriched SSO-based improved information gain, for detecting zero-day attacks, which may have applications in ransomware detection. | Limited validation of the proposed model in ransomware detection scenarios. |
| [20] | To explore the concept of securability for combining security and reliability of critical infrastructures, including ransomware protection measures. | Theoretical. |

However, to optimise its utility, the dataset required deduplication and transformation to ensure data integrity and compatibility with analytical frameworks (Figure 1).

**FEATURE ENGINEERING USING THE UGRANSOME DATASET**
During the feature engineering process, the UGRansome dataset underwent refinement by eliminating three irrelevant attributes (Flag, Port, and USD), resulting in the creation of the *UGRansome2024* dataset. This optimization was aimed at reducing dimensionality to accelerate computation and improve the accuracy of the machine learning model experiment. Additionally, certain remaining columns, such as Netflow Bytes, Expended, and Seed addresses, were renamed to fit within column space constraints. In the data labelling phase, adjustments were made to the target variable of the UGRansome dataset, as well as to the seed and expended addresses (Figure 2). The following modifications were made to align these variables with aspects of blockchain technology pertinent to crypto mining [21]:

- In the original dataset, the target variable "*Signature (S)*" has been renamed to "*Segwit*," reflecting its association with segment identification [21].

- The synthetic *signature (SS)* has been rebranded as "*Lightning Network*," [21] aligning it with the specific transaction type it represents.
- The *anomaly (A)* label has been updated to "Unusual Transaction," denoting transactions that deviate from typical patterns [21].

Furthermore, distinctions have been made among various transaction types, such as "*Pay-to-Public-Key-Hash Address (P2PKH)*," "*Pay-to-Script-Hash Addresses (P2SH)*," and "*Pay-to-Witness-Public-Key-Hash (P2WPKH),*" [21] to provide clarity and specificity within the dataset (Figure 3 and Figure 7).

(a) Features in the UGRansome Dataset

(b) Features in the UGRansome2024 Dataset

**Figure 3**: The UGRansome Dataset vs. the UGRansome2024

**RANSOMWARE DETECTION USING RANDOM FOREST**

Random Forest is a versatile machine learning algorithm that is used for both classification and regression tasks [2]. It operates by constructing a multitude of decision trees during training time and outputs the mode of the classes (classification) or the mean prediction (regression) of the individual trees [2]. We present how Random Forest works:

- **Random Sampling**: Random subsets of the training data are sampled (with replacement) to train each decision tree. This process is known as bootstrap sampling [2].

- **Feature Randomness**: At each node of the decision tree, a random subset of features is considered for splitting. This helps to decorrelate the trees and reduce overfitting [2].

- **Decision Tree Construction**: Each decision tree is constructed independently based on the random subsets of data and features. The trees are typically grown to maximum depth without pruning [2].

- **Voting or Averaging**: For classification tasks, the mode (most frequent class) of the predictions from individual trees is taken as the final prediction [2]. For regression tasks, the mean prediction of all trees is computed [2].

Random Forest offers several advantages:

- It is less prone to overfitting compared to individual decision trees.
- It can handle large datasets with high dimensionality.
- It provides estimates of feature importance, which can be useful for feature selection.

We have used Random Forest as a powerful algorithm in machine learning to classify ransomware using the UGRansome2024 dataset.

**RANDOM FOREST RESULTS USING UGRANSOME2024 DATA**

The UGRansome2024 dataset underwent encoding using the Python label encoder function, as illustrated in Figure 4. Figure 5 showcases the feature relevance determined by the Random Forest algorithm, while Figure 6 presents the confusion matrix generated by the same algorithm. Furthermore, the ransomware classification outcomes achieved by the Random Forest are depicted in Figure 7.

Notably, the Random Forest exhibited an impressive classification accuracy of 96%, as highlighted in Figure 6. Through this analysis, it becomes evident that the Random Forest model effectively identified Unusual Ransomware Transactions characterised by significant bitcoin (BTC) amounts, as depicted in Figure 7. Noteworthy findings indicate that Encrypt Decrypt Algorithms (EDA) and Globe ransomware incidents resulted in the highest financial impact, quantified in BTC, as evidenced in Figure 8.

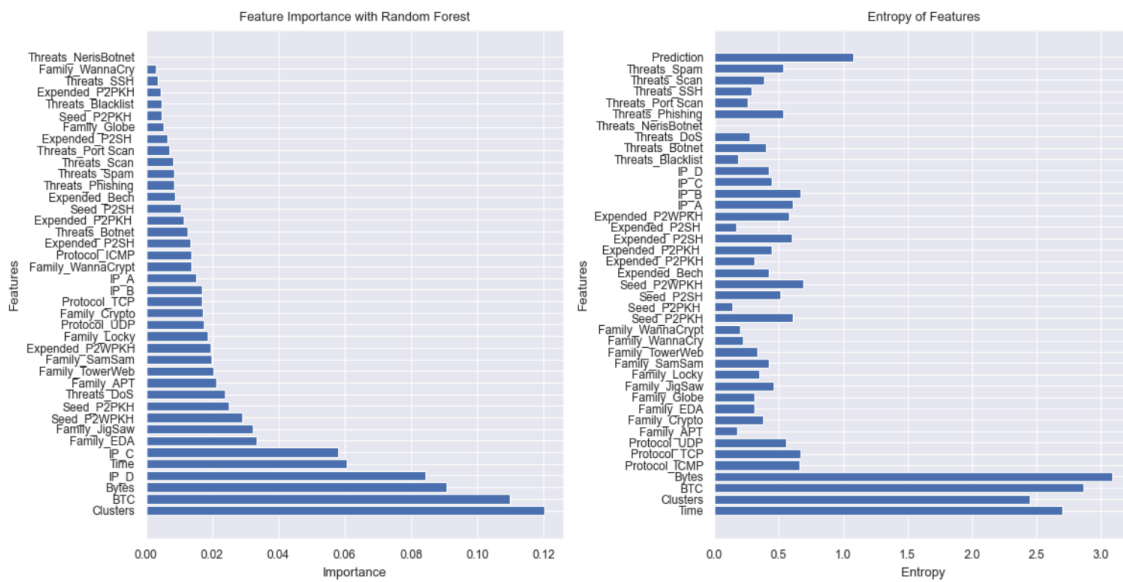

**Figure 4:** Encoded UGRansome2024 Data

**Figure 5**: Random Forest Results

```
Accuracy of Random Forest : 0.964
Classification report of Random Forest :
              precision    recall  f1-score   support

           0       0.98      0.94      0.96        53
           1       0.97      0.96      0.96        94
           2       0.95      0.99      0.97        78

    accuracy                           0.96       225
   macro avg       0.97      0.96      0.96       225
weighted avg       0.96      0.96      0.96       225

Confusion Matrix of Random Forest :
[[50  2  1]
 [ 1 90  3]
 [ 0  1 77]]
```

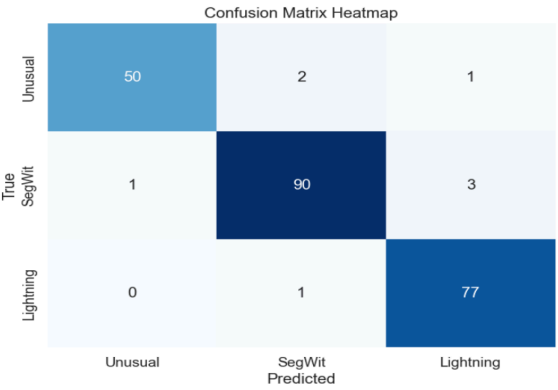

**Figure 6**: The Confusion Matrix of the Random Forest

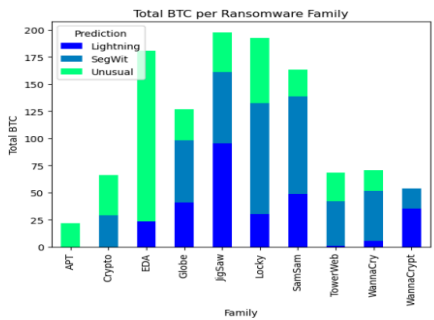

(a) Financial Impact per Ransomware

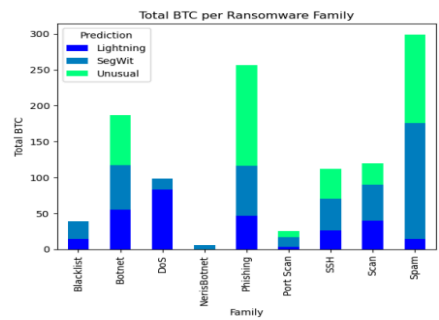

(b) Financial Impact per Malware

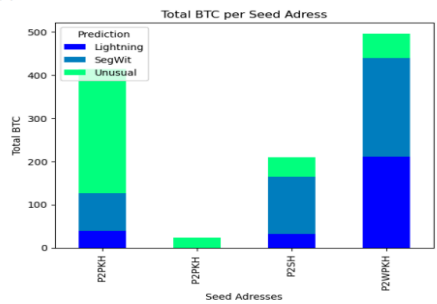

(c) Financial Impact per Seed Addresses

**Figure 7**: Ransomware Classification

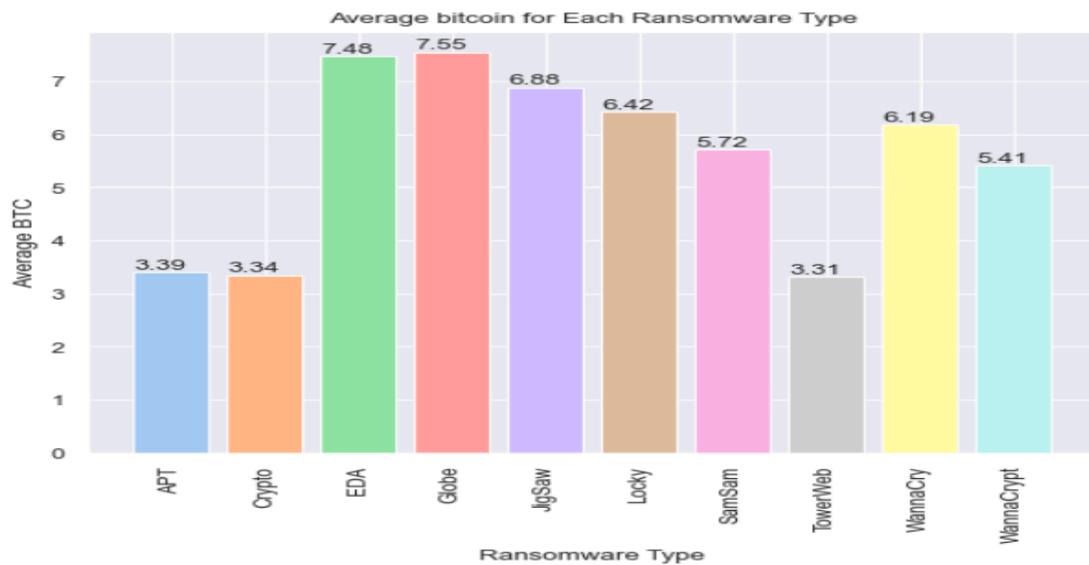

**Figure 8**: Financial Damage per Ransomware

**SIGNIFICANCE OF THE FINDINGS**
The findings from the analysis of the UGRansome2024 dataset using the Random Forest algorithm hold significant implications for ransomware detection and cybersecurity. Achieving a classification accuracy of 96% underscores the efficacy of machine learning techniques in identifying ransomware attacks, particularly those involving Unusual Transaction patterns. The identification of EDA and Globe ransomware as having the highest financial impact highlights the importance of understanding specific ransomware variants and their associated risks.

**APPLICABILITY IN REAL-LIFE**
The insights gained from this study can be applied in real-world cybersecurity scenarios to enhance ransomware detection and mitigation strategies. By leveraging machine learning algorithms such as Random Forest, organisations and cybersecurity professionals can develop more robust defence mechanisms to protect against ransomware attacks. Understanding the financial impact of different ransomware variants can also inform decision-making processes regarding resource allocation and risk management.

**FUTURE STUDIES**
Future research in this field could focus on expanding the UGRansome dataset to include a broader range of ransomware variants and attack scenarios. Additionally, exploring alternative machine learning algorithms and techniques for ransomware detection could provide further insights into improving detection accuracy and efficiency. Further investigation into the underlying factors contributing to the financial impact of ransomware incidents could also yield valuable insights for risk assessment and mitigation strategies.

**LIMITATIONS**
It is essential to acknowledge the limitations of this study, including the reliance on a specific dataset and machine learning approach. The generalizability of the findings may be limited by the scope and characteristics of the dataset used. Additionally, the effectiveness of the Random Forest algorithm may vary depending on the nature and complexity of ransomware attacks encountered in real-world scenarios.

**RECOMMENDATIONS**
Based on the findings of this study, it is recommended that organisations invest in robust cybersecurity measures, including the implementation of machine learning-based ransomware detection systems. Continuous monitoring and updating of datasets and detection algorithms are crucial to adapt to evolving ransomware threats effectively. Collaboration between cybersecurity researchers, practitioners, and industry stakeholders is also essential to share knowledge and best practices in ransomware detection and mitigation.

**CONCLUSION**
In conclusion, the analysis of the UGRansome2024 dataset using the Random Forest algorithm has provided valuable insights into ransomware detection and classification. The high classification accuracy achieved demonstrates the potential of machine learning in combating ransomware threats. By understanding the financial impact of different ransomware variants and employing advanced detection techniques, organisations can enhance their cybersecurity posture and better protect against ransomware Attacks.

**DATASET AND CODE AVAILABILITY**
Mike Wa Nkongolo 2023, UGRansome Dataset, Kaggle, viewed 11 April 2024, https://www.kaggle.com/dsv/7172543. DOI: 10.34740/KAGGLE/DSV/7172543